# Superfluid density in $Bi_2Sr_2CaCu_2O_{8+x}$ from optimal doping to severe underdoping and its implications

Jie Yong,[1,*] M. Hinton[1], A. McCray,[1] M. Randeria[1], M. Naamneh,[2] A. Kanigel,[2] and T.R. Lemberger[1]

[1]Department of Physics, The Ohio State University, Columbus, OH, USA
[2]Department of Physics, Technion - Israel Institute of Technology, Haifa 32000, Israel

[*]E-mail: yong.13@osu.edu

Due to their proximity to an antiferromagnetic phase and to the mysterious pseudogap, underdoped cuprates have attracted great interest in the high $T_c$ community for many years. A central issue concerns the role of quantum and thermal fluctuations of the phase of the superconducting order parameter. The evolution of superfluid density $n_s$ with temperature and doping is a powerful probe of this physics. Here, we report superfluid density measurements on underdoped $Bi_2Sr_2CaCu_2O_{8+x}$ (Bi-2212) films at much lower dopings than have been achieved previously, and with excellent control on doping level - $T_c$ ranges from $T_{c,min}$ ~ 6K to $T_{c,max}$ ~ 80K in steps of about 5K. Most famous studies on Bi-2212 like angle-resolved photoemission and scanning probe microscopy are surface-sensitive while superfluid density measurements are bulk-sensitive. We find that strong two-dimensional quantum fluctuations are evident in the observed linear scaling of $T_c$ with $n_s(0)$ when $T_c$ is below about 45 K, which contrasts with three-dimensional quantum fluctuations evident in the square root scaling, $T_c \propto \sqrt{n_s(0)}$, seen in the much less anisotropic cuprate, $YBa_2Cu_3O_{7-\delta}$ (YBCO)[1,2]. On the other hand, consistent with YBCO[1-3], $n_s(T)$ in severely underdoped Bi-2212 loses its strong downward curvature near $T_c$, becoming quasi-linear without any obvious critical behavior near $T_c$. We argue that the quasi-linear T dependence arises from thermal phase fluctuations, although the current theory needs modification in order to understand some features.

Early measurements[4] on moderately underdoped cuprate superconductors suggested a linear proportionality between superfluid density $n_s(0)$ and transition temperature $T_c$. The scaling was ascribed to the influence of thermal phase fluctuations[5] amplified by the low superfluid density $n_s(0)$ of underdoped cuprates, without reference to a mechanism for suppression of superfluid density with underdoping. On this basis, the linear scaling was expected to extend down to the superconductor to insulator quantum phase transition at very low doping.

Contrary to expectations, measurements on severely underdoped crystals[1,3] and films[2] of $YBa_2Cu_3O_{7-\delta}$ (YBCO) have revealed a sublinear relationship, $T_c \propto n_s(T)^\alpha$ with α ~½, indicating that underdoping suppresses $n_s$ at T = 0 by generating 3D quantum phase fluctuations (with a dynamical critical exponent, $z_Q \approx 1$)[6,7]. Consequently, $T_c$ and $n_s(0)$ disappear at a 3D quantum critical point (QCP). For samples near the QCP, quantum and thermal phase fluctuations, acting in concert, drive $n_s(T)$ to zero at a temperature that scales with a power of $n_s(0)$. This physics is clear in the case of 2-unit-cell thick Ca-doped YBCO films[8], where both 2D quantum critical scaling, $T_c \propto n_s(0)$, and 2D thermal critical behavior, namely, an abrupt Kosterlitz-Thouless-like drop in $n_s(T)$ near $T_c$, are both observed. The 3D nature of critical fluctuations in bulk YBCO means that there is significant coupling between neighboring $CuO_2$ bilayers. Another interesting phenomenon is that evidence for critical fluctuations near $T_c$ disappears from severely underdoped YBCO crystals[1,3] and films[2]. That is, the strong downward curvature seen in $n_s(T)$ near $T_c$ in moderately underdoped YBCO disappears when doping is severe enough that $T_c/T_{c,max} \lesssim \frac{1}{2}$. Within the theory of quantum critical phenomena, it is entirely possible for the thermal critical region to become too narrow to detect[7], another indication of strong quantum fluctuations.

We emphasize that the c-axis vs. ab-plane anisotropy of Bi-2212 is much larger than for YBCO, so fluctuations could be qualitatively different. Near optimal doping, $Bi_2Sr_2CaCu_2O_{8+x}$ (Bi-2212) has a resistivity anisotropy[9,10] of at least $10^5$ compared to only about 100 in YBCO[11].

Our samples are Bi-2212 films grown by pulsed laser deposition (PLD) onto MgO substrates at OSU and by sputtering onto $LaAlO_3$ substrates at Technion. Targets are stoichiometric $Bi_2Sr_2CaCu_2O_{8+x}$. Hole doping is tuned by oxygen pressure during deposition and postannealing. All the films are epitaxial and phase pure as indicated by XRD measurements. FWHM values of rocking curves on film peaks are always 0.3 to 0.4 degrees. Severely underdoped films tend to have slightly larger out-of-plane lattice constants than moderately underdoped films, which agree with YBCO data[12]. All the films are about 100 nm thick. Transition temperature varies from $T_{c,max}$~80K for near optimal doped films to $T_{c,min}$ ~ 6K for severely underdoped ones.

The great advantage of studying films is that we are able to underdope them by reducing oxygen concentration to much lower $T_c$'s than is possible in bulk. The substrate probably provides mechanical stability. As discussed below, our films agree with bulk at dopings where the bulk data exist, suggesting that our results are characteristic of Bi-2212 and are insensitive to whatever flaws exist in films. To back up this notion, PLD and sputtered films with similar $T_c$'s have similar superfluid densities both in magnitude and T-dependence, even though these films likely have different types of structural and chemical flaws, given the differences in deposition process and in substrate material.

Figure 1 shows $\lambda^{-2}(T)$ of many Bi-2212 films. Red/pink curves are sputtered films and blue/green curves are PLD films. Peaks in $\sigma_1(T)$ near $T_c$ indicate all the films have decently sharp transitions, with transition widths of about 2 K for the most strongly underdoped samples. Sputtered and PLD films agree well with each other, as seen from pairs of films with $T_c$'s near 53 K and 37 K. Also, note the reproducibility of two PLD films that happen to have the same $T_c$ of 28 K. Our moderately underdoped films show similar temperature dependencies to those of optimally-doped ($T_c \approx 90$ K) and moderately underdoped ($T_c \approx 80$ K, 65 K) $Bi_{2.15}Sr_{1.85}CaCu_2O_{8+x}$ powders[13], although the downturn near $T_c$ is not as dramatic in powders. We will see below that scaling of $T_c$ with $n_s(0)$ is the same for films and powders at moderate underdoping. Thus, our films are of comparable quality to bulk materials.

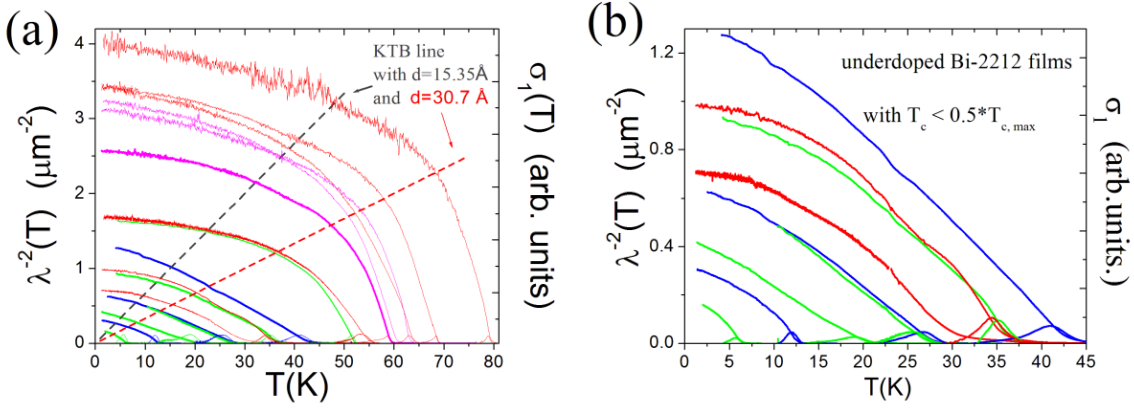

**Figure 1: T-dependence of superfluid density for underdoped Bi-2212 films (PLD films in green/blue; sputtered films in red/pink).** (a) full range of dopings. Intersection of upper (lower) dashed line with $\lambda^{-2}(T)$ is where Kosterlitz-Thouless theory predicts a downturn in $\lambda^{-2}$ due to vortex-antivortex unbinding for d=15.35 Å (1 $CuO_2$ bilayer) [d=30.7 Å (2 $CuO_2$ bilayers)]. (b) films with $T_c$ < 45 K. The widths of peaks in $\sigma_1$ near $T_c$ set an upper limit on the spatial inhomogeneity of $T_c$.

Clear features for moderately underdoped films, $T_c$ > 50 K, are the weak linear T-dependence at low-T and sharp downturn near $T_c$. Given the extremely anisotropic nature of Bi-2212, we associate this downturn with quasi-2D thermal phase fluctuations. For 2D superconductors, Kosterlitz-Thouless theory[14] predicts a downturn where $k_BT$, ($k_B$ = Boltzmann's constant) is about equal to the energy required to create a vortex-antivortex pair, i.e., the transition occurs at the temperature where: $\lambda^{-2}(T) = (2\mu_0/\pi d\hbar R_Q)k_BT$. The quantum resistance $R_Q \equiv \hbar/4e^2 \approx 1$ k$\Omega$, and $d$ is the effective 2D thickness. The two dashed lines in Fig. 1 represent the rhs of this equation assuming that $d$ = 15.35 Å (1 $CuO_2$ bilayer) and $d$ = 30.7 Å (2 $CuO_2$ bilayers). Intersection of the latter line with the observed $\lambda^{-2}(T)$ is consistent with what is seen in 2 unit cell thick YBCO films[8] and conventional 2D superconducting films[15]. Thus, we conclude that the extreme anisotropy of Bi-2212 brings the effective 2D layer thickness down to 2 bilayers as regards classical thermal fluctuations.

It is odd that the sharp downturn in $\lambda^{-2}(T)$ diminishes with more severe underdoping, where phase fluctuations should be enhanced. Also odd is the abruptness of the downturn's disappearance, which occurs when $T_c$ drops below about 48K. As seen below, this is the same place where scaling of $T_c$ with $\lambda^{-2}(0)$ becomes linear. As for the former, Franz and Iyengar[7] have explained that the thermal critical regime can narrow dramatically for samples near a QCP, thereby accounting for the apparent disappearance of downward curvature near $T_c$. If the downturn in $\lambda^{-2}(T)$ for moderate underdoping is due to critical thermal phase fluctuations, and its apparent disappearance is due to quantum fluctuations near a QCP, it is natural to suppose that quantum and thermal phase fluctuations conspire to dominate the T-dependence of $\lambda^{-2}(T)$ for severely underdoped films. The following analysis shows that this notion is reasonable, with some caveats.

Classical thermal phase fluctuations are expected to suppress superfluid density by a factor of $(1 - k_B T/4J)$ in square arrays of superconducting grains coupled by Josephson energy J, as long as the suppression is small[16]. For continuous films, the effective coupling energy J is proportional to the sheet superfluid density: $J(T) = \hbar R_Q d \lambda^{-2}(T)/\mu_0$, where $d$ is the effective 2D thickness. Thus, the low-T behaviour would be:

$$\lambda^{-2}(T) \approx \lambda^{-2}(0)[1 - k_B T/4J(0)] = \lambda^{-2}(0) - \mu_0 k_B T/4\hbar R_Q d.$$

The low-T slope of $\lambda^{-2}(T)$ does not depend on $\lambda^{-2}(0)$ in this model. Quantum effects would change linear to quadratic behaviour below some cutoff temperature[17], but we set that physics aside. We see in the measured $\lambda^{-2}(T)$ for films with $T_c < 48$ K that the slopes of the quasilinear data are, indeed, roughly independent of $\lambda^{-2}(0)$. Quantitatively, the slope should be -0.013 $\mu m^{-2}$/K, assuming $d = 30.5$ Å, which is close to the measured slopes. Of course, we have used a low-T expression to analyse data up to $T_c$. Further theoretical work is needed to determine whether quantum critical fluctuations would repair this discrepancy.

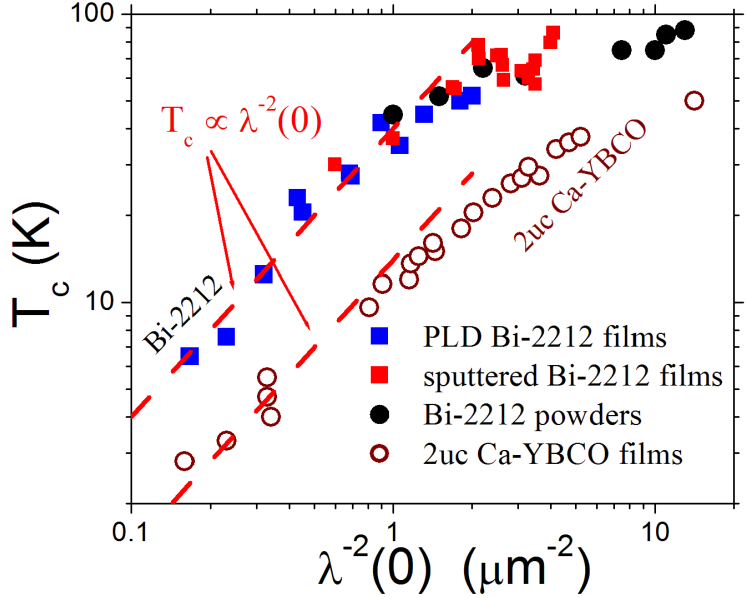

**Figure.2: Scaling between $T_c$ and $\lambda^{-2}(0)$ for underdoped Bi-2212 from various growth methods (PLD – blue squares; sputtered – red squares; powders – black circles).** Data for 2-unit-cell thick Ca-YBCO films (open red circles) are plotted in comparison. Two dashed lines show a linear relationship between $T_c$ and $\lambda^{-2}(0)$.

Although anomalous, the quasi-linear T-dependence of $\lambda^{-2}$ at severe underdoping is also seen in severely underdoped YBCO crystals[1,3] and films[2]. Apparently this linear behaviour is robust against anisotropy and disorder. It may be regarded as a universal phenomenon for severely underdoped cuprates, and it therefore provides important guidance to theory.

Having argued for strong quantum critical fluctuations on the basis of apparent absence of thermal critical behaviour, we turn to a second key indicator, namely, power-law scaling of $T_c$ with $\lambda^{-2}(0)$. This scaling, if present, is sensitive to dimensionality. If quantum fluctuations are 2D, then theory of quantum critical scaling[6] requires that $T_c \sim [\lambda^{-2}(0)]^\alpha$ where $\alpha = 1$, regardless of details.

Figure 2 shows a log-log plot of $T_c$ vs. $\lambda^{-2}(0)$ for Bi-2212 films and powders[13] and, for comparison with a truly 2D system, for 2-unit-cell thick YBCO films. Note that films agree with powders to the lowest dopings achieved in powders, and that sputtered and PLD films agree with each other. Clearly, scaling is linear at the lowest dopings, indicating quantum fluctuations are 2D in Bi-2212 due to its extremely high anisotropy. In YBCO scaling is square-root, indicating 3D quantum fluctuations[1,3]. The quasi-2D nature of Bi-2212 is apparent in other measurements, e.g., $T_c$ is unaffected by intercalation of various materials into BiO bilayers[19], reducing coupling between adjacent $CuO_2$ bilayers. Setting aside theory, the semi-quantitative similarity between Bi-2212 and 2-unit-cell thick YBCO films (Fig.2) indicates similar fluctuation physics. The main qualitative difference is that Kosterlitz-Thouless physics is seen in the T-dependence of $\lambda^{-2}(T)$ for severely underdoped 2D YBCO films but not in Bi-2212. Apparently the very weak interlayer coupling in Bi-2212 is enough to permit quantum fluctuations to narrow the critical thermal region beyond visibility.

Finally, we note that it is surprising that linear scaling begins so abruptly as underdoping brings $T_c$ below about 48 K, the same point where the T-dependence of $\lambda^{-2}$ changes to quasi-linear.

`

**Table I: A comparison of this work with superfluid density measurements on other underdoped cuprate systems.**

| System | thick YBCO films or crystals | 2 unit cells of YBCO or LSCO | thick Bi-2212 films |
|---|---|---|---|
| Anisotropy | moderate ($\rho_c/\rho_{ab} \sim 10^2$) | 2-D by construction | extreme ($\rho_c/\rho_{ab} \sim 10^5$) |
| Near optimal doping | 3D XY-behaviour (crystals) | KT Transition | KT-like Transition |
| Severely underdoped | $\lambda^{-2}$ linear up to $T_c$, no critical behavior | KT Transition | $\lambda^{-2}$ linear up to $T_c$, no critical behavior |
| Scaling: $T_c \propto [\lambda^{-2}(0)]^\alpha$ | $\alpha \sim 1/2$, 3D Quantum critical fluctuations | $\alpha \sim 1$, 2D Quantum critical fluctuations | $\alpha \sim 1$, 2D Quantum critical fluctuations |
| When add intercalation layer between $CuO_2$ bilayers | $T_c$ changes | N.A. | Both $T_c$ and sheet superfluid density per $CuO_2$ bilayer stay the same |
| References | 1-3, 11,18, 20 | 8, 21 | 9, 10, 19 and this work |

To put our work into context, we put our results together with previous work in Table I. This table summarizes the similarities and differences from T-dependent superfluid density measurements in three different systems. A careful comparison can help us have a better understanding on the anisotropy nature of cuprates.

Acknowledgements: Work at OSU was supported by DOE-Basic Energy Sciences through grant FG02-08ER46533 (JY, MH, AM, TRL), by the OSU Dept. of Physics (AM), and by NSF-DMR 0706203 (MR). We acknowledge useful discussions with Douglas Bonn.

# Methods

Superfluid Density Measurements

Superfluid densities are measured by a two-coil mutual inductance apparatus[22, 23]. The film is sandwiched between two coils, and the mutual inductance between these two coils is measured at a frequency $\omega/2\pi$ = 50 kHz. The measurement actually determines the sheet conductivity, $Y \equiv (\sigma_1 + i\sigma_2)d$, with $d$ being the superconducting film thickness and $\sigma$ being the conductivity. Given a measured film thickness, $\sigma$ is calculated as: $\sigma = Y/d$. The imaginary part, $\sigma_2$, yields the superfluid density through: $\omega\sigma_2 \equiv n_s e^2/m$, which is proportional to the inverse penetration depth squared: $\lambda^{-2}(T) \equiv \mu_0 \omega \sigma_2(T)$, where $\mu_0$ is the permeability of vacuum. As is customary, we refer to $\lambda^{-2}$ as the superfluid density. The dissipative part of the conductivity, $\sigma_1(T)$, has a peak near $T_c$, whose width provides an upper limit on the spatial inhomogeneity of $T_c$ over the 10 mm$^2$ area probed by the measurement. Data are taken continuously as the sample slowly warms up so as to yield the hard-to-measure absolute value of $\lambda$ and its T-dependence.